\documentclass[12pt,epsf]{revtex4}
\usepackage{graphicx} 

\begin{document}

\Large
\title{\bf Fulde-Ferrell-Larkin-Ovchinnikov - like state in 
           Ferromagnet - Superconductor Proximity System}

\vspace{0.5cm}
\large
\author{B. L. Gy\"{o}rffy$^{1,2}$, M. Krawiec$^3$, J. F. Annett$^1$}

\vspace{0.5cm}
\normalsize
\affiliation{$^1$ H. H. Wills Physics Laboratory, University of Bristol, 
             Tyndall Ave., Bristol BS8 1TL, UK \\
             $^2$ Centre for Computational Materials Science, TU Wien, 
	     Gertreidemarkt 9/134, A-1060 Wien,Austria \\
	     $^3$ Institute of Physics and Nanotechnology Center, Maria 
	     Curie-Sk\l odowska University, Pl. Marii Curie-Sk\l odowskiej 1, 
	     20-031 Lublin, Poland}


\date{\today}


\vspace{0.5cm}
\begin{abstract}
We discuss some properties of the ferromagnet - superconductor proximity 
system. In particular, the emphasis is put on the physics of the
Fulde-Ferrell-Larkin-Ovchinnikov ($FFLO$) like state. In addition to Andreev
reflections it features a number of unusual thermodynamic and transport 
properties, like: oscillatory behavior of the pairing amplitude, density of 
states and superconducting transition temperature as a function of the 
ferromagnet thickness. Surprisingly, under certain conditions spontaneous spin 
polarized current is generated in the ground state of such a system. We provide 
some informations regarding experimental observations of this exotic state.
\end{abstract}

\maketitle


\section{\label{sec1} Introduction}


When a normal non-magnetic metal is connected to a superconductor it acquires
superconducting properties, like non-zero pairing amplitude. This effect, known 
as the proximity effect \cite{Lambert}, has extensively been studied for almost 
half a century. It is rather well understood by now in terms of Andreev 
reflections \cite{Andreev}, according to which an impinging electron (with 
energy less than superconducting gap) on the normal metal ($NM$) / 
superconductor ($SC$) interface is reflected back as a hole and the Cooper pair 
is created in superconductor. From the point of view of Andreev reflections the 
proximity effect can be regarded as a non-zero density of the Andreev 
correlated electron - hole pairs on the normal metal side of the interface.

When a normal metal is replaced by a ferromagnet ($FM$), another energy scale
enters problem, namely the exchange splitting which is related to the spin
polarization of the electrons. Such $FM$/$SC$ hybrid structures are important 
from the scientific point of view, as they allow the study of the interplay 
between ferromagnetism and superconductivity \cite{Berk} as well as of device
applications in such areas of technology as magnetoelectronics \cite{Bauer} or
quantum computing \cite{Blatter}.

It is widely accepted that ferromagnetism and superconductivity are two
antagonistic phenomena, so one could expect that the proximity effect in
$FM$/$SC$ system should be suppressed. Indeed, the one can argue that in
ferromagnet there are different numbers of spin-up (majority) $n_{\uparrow}$
and spin-down (minority) $n_{\downarrow}$ conduction channels, and due to the
fact that incident and reflected particles occupy different spin bands, only
a fraction $n_{\downarrow}$/$n_{\uparrow}$ of majority particles can be
Andreev reflected \cite{deJong}. 

On the other hand if an exchange field acts on the Cooper pairs, one would 
expect that either it is too weak to break the pair, or it suppresses completely
superconductivity. However when a Cooper pair is subjected to the exchange 
field, it acquires a finite momentum and for certain values of the exchange 
splitting a new superconducting state is realized, known as 
{\it Fulde - Ferrell - Larkin - Ovchinnikov} ($FFLO$) state \cite{FF,LO}.
Interestingly such state features a spatially dependent order parameter
corresponding to the non-zero center of mass motion of the Cooper pairs. 
This state features in non-zero spin polarization, almost normal tunneling
characteristics and almost normal Sommerfeld specific heat ratio, anisotropic
electrodynamic properties. Unfortunately the bulk state is very sensitive to 
the impurities and shape of the Fermi surface. Another novel feature of this 
state is a current flowing in the ground state. The unpaired electrons tend to 
congregate at one portion of the Fermi surface so a quasiparticle current is 
produced. In order to satisfy the Bloch theorem: no current in the ground 
state, a supercurrent, generated by the nonzero value of the pairing momentum, 
flows in opposite direction, and the total current is zero. 

Similar oscillations of the pairing amplitude have been predicted 
\cite{Buzdin}-\cite{Proshin} in ferromagnet/superconductor proximity systems. 
It turns out that these oscillations are responsible for the oscillatory
behavior of the $SC$ critical temperature $T_c$, first experimentally observed 
by Wong {\it et al.} \cite{Wong}, and the density of states \cite{Kontos} as 
the thickness of the $FM$ slab is varied. In fact, the oscillations of the 
$T_c$ in $FM$/$SC$ multilayers can be also explained in terms of the effective 
$\pi$-junction behavior \cite{Radovic}. It was shown that at specific $FM$
thickness the Josephson coupling between two $SC$ layers can lead to a junction
with an intrinsic phase (of the order parameter) difference 
$\delta \varphi = \pi$, which exhibits a higher $T_c$ than the ordinary one
($\delta \varphi = 0$). The $\pi$-junction effect has been originally proposed 
by Bulaevskii {\it et al.} \cite{Bulaevskii} to arise in the tunnel barriers 
containing magnetic impurities. It was also suggested that the $\pi$-junction can be 
realized in high-$T_c$ superconducting weak links \cite{Sigrist}, where the 
$SC$ order parameter changes its sign under $\pi/2$ rotation. This has 
tremendous consequences as it leads to many important effects
\cite{pi_junct,Sigrist_1}, like: the zero energy Andreev states, zero-bias 
conductance peaks, large Josephson current, time reversal symmetry breaking, 
paramagnetic Meissner effect and spontaneously generated currents.

From the point of view of the present paper the important issue is the
formation of the Andreev bound states in $FM$/$SC$ proximity system. The 
Andreev states arise due to the fact that the quasiparticles of the ferromagnet 
participating in the Andreev reflections move along closed orbits. Such states 
have been first studied by de Gennes and Saint-James \cite{deGennes} in the 
insulator/normal metal/superconductor ($I$/$NM$/$SC$) trilayer. The energies of 
these states are always smaller than the $SC$ gap $\Delta$ and symmetrically 
positioned around the Fermi level. They strongly depend on the geometry of the 
system as well as on the properties of the interfaces. In high-$T_c$ ($d$-wave) 
superconductors, these states can be shifted to zero energy, due to the 
specific form of the symmetry of the order parameter \cite{Hu}, thus indicating 
$\pi$-junction behavior in the system. Naturally, such Andreev states can also 
arise in the $I$/$FM$/$SC$ heterostructures. Moreover, it is possible to shift 
the energies of these states by changing the exchange splitting, as was first 
demonstrated by Kuplevakhskii \& Fal'ko \cite{Kupl}. In turn, by properly 
adjusting the exchange splitting the position of the Andreev bound states can 
be moved to the Fermi energy. The system under such circumstances behaves like 
that being in the $\pi$-junction phase as the spontaneous current is generated 
\cite{KGA}. 

Some of our results have already been published \cite{KGA}-\cite{KGA_2}. Here 
we wish to present a more detailed study of the $FM$/$SC$ proximity system in 
terms of $FFLO$ physics. In some situations the ground state of $FM$/$SC$ 
structures has properties of both the $FFLO$ and the $\pi$-junction, leading 
to various interesting and unexpected phenomena. 

The paper is organized as follows: In Sec. \ref{sec2} the simple model which 
allows for self-consistent description of the $FM$/$SC$ heterostructure is 
introduced. In Sec. \ref{sec3} the nature of the Andreev bound states in the 
ferromagnet is discussed. The spontaneously generated current and corresponding 
magnetic field in the ground state are studied in the Sec. \ref{sec4}. In 
Sec. \ref{sec5} show some transport properties of the system, in the 
Sec. \ref{sec6} we compare our system to usual $FFLO$ state, and finally, we 
conclude in Sec. \ref{sec6}.


\section{\label{sec2} Model and theory}


To study the properties of $FM$/$SC$ system we have adopted the $2D$ Hubbard
model featuring the exchange splitting in the ferromagnet and an electron -
electron attraction in superconductor. The Hamiltonian is:
\begin{eqnarray}
 H = 
 \sum_{ij\sigma} \left[ t_{ij} + \left( \frac{1}{2} E_{ex} \sigma - \mu \right) 
 \delta_{ij} \right] c^+_{i\sigma} c_{j\sigma} + 
 \frac{1}{2} \sum_{i\sigma} U_i n_{i\sigma} n_{i-\sigma}
 \label{Hamiltonian}
\end{eqnarray}
where in the presence of a vector potential $\vec{A}(\vec{r})$, the hopping
integral is given by $t_{ij} = - t e^{-i e \int_{\vec{r}_i}^{\vec{r}_{j}}
\vec{A}(\vec{r}) \cdot d\vec{r}}$ for nearest neighbor lattice sites, whose
positions are $\vec{r}_i$ and $\vec{r}_j$, and zero otherwise. The exchange
splitting $E_{ex}$ is only non-zero on the $FM$ side, unlike as $U_i$ (electron
- electron attraction) being non-zero only in $SC$. $\mu$ is the chemical
potential, $c^+_{i\sigma}$, ($c_{i\sigma}$) are the usual electron
creation (annihilation) operators and
$\hat n_{i\sigma} = c^+_{i\sigma} c_{i\sigma}$.

In the following we shall work within Spin - Polarized - Hartree - Fock -
Gorkov ($SPHFG$) approximation \cite{KGA} assuming periodicity in the direction
parallel to the interface while working in a real space in the direction
perpendicular. Labeling the layers by integer $n$ and $m$ at each $k_y$ point
of the Brillouin zone we shall solve the following $SPHFG$ equation:
\begin{eqnarray}
 \sum_{m',\gamma,k_y} H^{\alpha\gamma}_{nm'}(\omega,k_y)
 G^{\gamma\beta}_{m'm}(\omega,k_y) =
 \delta_{nm} \delta_{\alpha\beta}
 \label{HFG}
\end{eqnarray}
where the only non-zero elements are:
$H^{11}_{nm}$ and $H^{22}_{nm} = (\omega - \frac{1}{2} \sigma E_{ex} \pm \mu
 \pm  t cos(k_y \mp eA(n)))\delta_{nm} \pm t \delta_{n,n+1}$ for the upper and
lower sign respectively,
$H^{33}_{nm} = H^{11}_{nm}$ and $H^{44}_{nm} = H^{22}_{nm}$ with $\sigma$
replaced by $-\sigma$ and
$H^{12}_{nm} = H^{21}_{nm} = - H^{34}_{nm} = - H^{43}_{nm} =
\Delta_n \delta_{nm}$
and $G^{\alpha\beta}_{nm}$ is corresponding Green's function ($GF$).

As usual, the self-consistency is assured by the relations determining the $FM$
($m_n$) and $SC$ ($\Delta_n$) order parameters, current 
($J_{y\uparrow (\downarrow)}(n)$) and the vector potential ($A_y(n)$)
respectively:
\begin{eqnarray}
 m_n = n_{n\uparrow} - n_{n\downarrow} =
 \frac{2}{\beta}
 \sum_{ky} \sum^{2N-1}_{\nu = 0}
 {\rm Re} \left\{
 (G^{11}_{nn}(\omega_{\nu},k_y) - G^{33}_{nn}(\omega_{\nu},k_y))
 e^{(2 \nu + 1) \pi i / 2 N}
 \right\}
 \label{m}
\end{eqnarray}
\begin{eqnarray}
 \Delta_n = U_n \sum_{k_y}
 \langle c_{n\downarrow}(k_y) c_{n\uparrow}(k_y) \rangle =
 \frac{2 U_n}{\beta}
 \sum_{ky} \sum^{2N-1}_{\nu = 0}
 {\rm Re} \left\{
 G^{12}_{nn}(\omega_{\nu},k_y) e^{(2 \nu + 1) \pi i / 2 N}
 \right\}
 \label{Delta}
\end{eqnarray}
\begin{eqnarray}
 J_{y\uparrow (\downarrow)}(n) =
 \frac{4 e t}{\beta} \sum_{k_y} sin(k_y - e A_y(n))
 \sum^{2N-1}_{\nu = 0}
 {\rm Re} \left\{
 G^{11(33)}_{nn}(\omega_{\nu},k_y) e^{(2 \nu + 1) \pi i / 2 N}
 \right\}
 \label{current}
\end{eqnarray}
\begin{eqnarray}
 A_y(n+1) - 2 A_y(n) + A_y(n-1) = - 4 \pi J_y(n)
 \label{Maxwell}
\end{eqnarray}
The details of the calculations can be found in \cite{KGA_1}.


\section{\label{sec3} Andreev bound states}


Before we discuss results of fully self-consistent calculations we would like 
to turn the attention to origin of Andreev bound states and take a look at
physics of them from the point of view of semiclassical approach. 

\begin{figure}[h]
\begin{center}
 \resizebox{9cm}{!}{
  \includegraphics{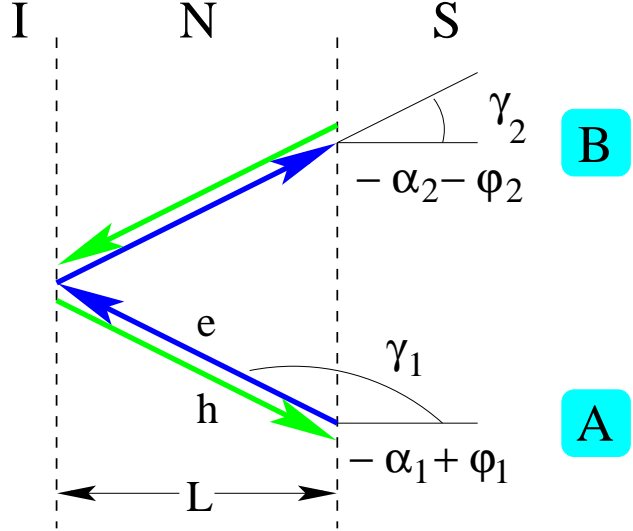}
}
\end{center}
 \caption{\label{Fig1} The example of the quasiparticle path 
          corresponding to the Andreev reflections, giving a bound state. The 
	  quasiparticle is trapped in the normal region because of normal 
	  reflection at the $I$/$NM$ surface and the Andreev reflection at the 
	  $NM$/$SC$ interface. The total phase accumulated during one cycle is 
	  equal: 
	  $-(\alpha_1 + \alpha_2) \pm (\varphi_1 - \varphi_2) + \beta(E)$.}
\end{figure}

From quasiclassical considerations, each bound state corresponds to 
quasiparticle moving along a family of closed trajectories \cite{Dunkan}. The 
energy of such bound state is determined by the Bohr-Sommerfeld quantization 
rules, according to which the total phase accumulated during one cycle has to 
be equal to multiples of $2\pi$. Interestingly, the bound states also emerge in 
the normal metal/superconductor ($NM$/$SC$) structures \cite{deGennes} due to 
the Andreev reflections \cite{Andreev}, according to which an incident electron 
is reflected back as a hole at the interface, and a Cooper pair is created in 
$SC$. Such states are built up from a combination of electron and hole wave
functions. The example of the closed quasiparticle trajectory, producing the
bound state, in an insulator/(normal metal)/superconductor $I$/$NM$/$SC$, is 
shown in the Fig. \ref{Fig1}. 
It consists of an electron $e$ segment, which includes a ordinary reflection 
at the $I$/$NM$ interface, and hole $h$ one, retracing backwards the electron 
trajectory. The total accumulated phase in this case consists of contribution
from Andreev reflections at point $A$: $-\alpha_1 + \varphi_1$ and $B$: 
$-\alpha_2 + \varphi_2$ as well as contribution from the propagation through 
the normal metal $\beta(E)$. $\alpha_{1(2)} = arccos{(E/|\Delta_0|)}$ is the
Andreev reflection phase shift, while $\varphi_{1(2)}$ is the phase of the 
$SC$ order parameter at point $A$ ($B$). 
$\beta(E) = 2 L (k_e - k_h) + \beta_0$ is the electron-hole dephasing factor 
and describes the phase acquired during the propagation through the normal 
region, where the first term corresponds to the ballistic motion and the 
second one to the reflection at the $I$/$NM$ surface. $L$ is the thickness of 
$NM$, and $k_e$ ($k_h$) is the electron (hole) wave vector. Thus the 
Bohr-Sommerfeld quantization condition is:
\begin{eqnarray}
 -(\alpha_1 + \alpha_2) \pm (\varphi_1 - \varphi_2) + \beta(E) = 2 n \pi
 \label{Bohr}
\end{eqnarray}
where the $\pm (\varphi_1 - \varphi_2)$ stands for the trajectories in the 
$\pm k_y$ (parallel to the interface) direction. 

If there is no phase difference between points $A$ and $B$ in the 
Fig. \ref{Fig1} (for example $NM$/$SC$ interface), the bound states always 
appear in pairs symmetrically positioned around the Fermi level because of the 
time reversal symmetry in the problem. Moreover, due to the fact that there is 
no difference between electrons and holes at the Fermi level 
($\beta(E = 0) = 0$), there is no $E = 0$ solution. In other words, the bound 
states always emerge at finite energies.

The situation is quite different if there is a phase difference 
$(\varphi_1 - \varphi_2)$ between points $A$ and $B$ (see Fig. \ref{Fig1}). 
The example can be the interfaces with $d$-wave superconductors oriented in 
the $(110)$ direction, where $(\varphi_1 - \varphi_2) = \pi$. In this case, 
due to the additional phase shift $\pi$ the bound states can emerge even at 
zero energy. Such zero-energy Andreev bound states, in the case of high-$T_c$ 
superconductors, have been predicted by Hu \cite{Hu} and are known as 
{\it zero-energy mid-gap states}. The presence of the Andreev bound states at 
zero energy features in many important effects, like zero-bias conductance 
peaks, $\pi$-junction behavior, anomalous temperature dependence of the 
critical Josephson current, paramagnetic Meissner effect, time reversal 
symmetry breaking and spontaneous interface currents \cite{pi_junct,Sigrist_1}. 

\begin{figure}[h]
\begin{center}
 \resizebox{9cm}{!}{
  \includegraphics{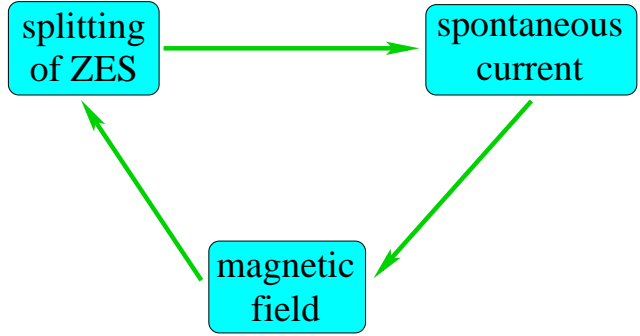}
}
\end{center}
 \caption{\label{Fig2} Generating of the spontaneous currents.}
\end{figure}

Although the zero-energy states ($ZES$) are likely to appear when the phase of 
the order parameter at the interface is not constant, the resulting density of
states at the Fermi energy is energetically unfavorable and any mechanism 
able to split these states will lower the energy of the system 
\cite{Sigrist_1,Fogelstrom}. On of these is the self-induced Doppler shift
\cite{Higashitani,pi_junct} $\delta = e v_F A$, where $A$ is a vector
potential. The situation is schematically depicted in the Fig. \ref{Fig2}. 
At low temperature ($T^* \approx (\xi_0/\lambda) T_c$, where $\lambda$ is the
penetration depth of the magnetic field) the splitting of the zero energy 
states produces a surface current. This current generates a magnetic field 
(screened by a supercurrent), which further splits $ZES$ due to the Doppler 
shift effect. The effect saturates when the magnetic energy is equal to the 
energy of the Doppler shifted $ZES$. 

Naturally, the Andreev bound states also arises in $I$/$FM$/$SC$ 
heterostructures \cite{Kupl,Kadigrobov,Zareyan,Belzig,Vecino,KGA_1,KGA_2,KGA_3}. 
More importantly, as it was first predicted by Kuplevakhskii \& Fal'ko 
\cite{Kupl}, it is possible to shift these states to zero energy by tuning 
the exchange splitting. So the crossing of the zero energy solution can be 
obtained either by changing the phase difference $(\varphi_1-\varphi_2)$ or by 
varying $FM$ coherence length (exchange field).

The properties of such bound states have been also studied fully 
quantum-mechanically within lattice models of the $FM$/$SC$ systems 
\cite{Vecino,KGA_1,KGA_2} and similar their behavior have been obtained.
Interestingly, it turns out, that as in the case of the high-$T_c$ structures 
\cite{Higashitani}, such zero energy Andreev states support spontaneous 
currents flowing in the ground state of the $FM$/$SC$ system 
\cite{KGA,KGA_1,KGA_2}. The mechanism of generating of such currents is the 
same, as earlier discussed, namely the self-induced Doppler shift. So in fact,
when the current flows, such one of the states will be twice shifted: once due
to the exchange (Zeeman) splitting, and the second time due to the Doppler 
shift. 

\begin{figure}[h]
\begin{center}
 \resizebox{9cm}{!}{
  \includegraphics{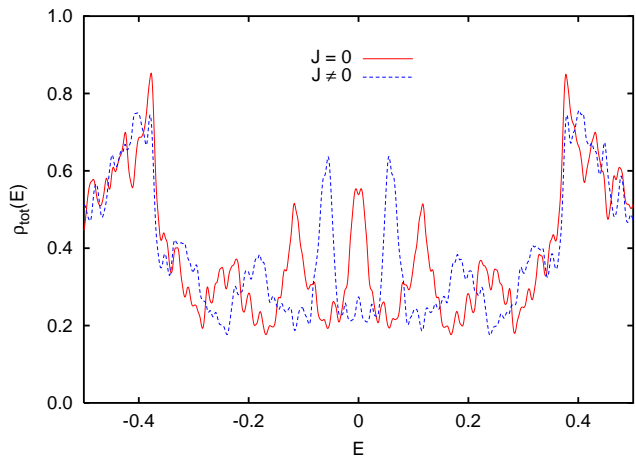}
}
\end{center}
 \caption{\label{Fig3} Doppler splitting of the zero-energy state. 
          From Ref. \cite{KGA_1}.}
\end{figure}

For energies less than superconducting gap, the only Andreev bound states will
contribute to the density of states $\rho(E)$. However, as it was mentioned, 
for fixed thickness and exchange splitting, there will be Andreev bound states 
at different energies, for different angles of particle incidence 
($\gamma_2$ in the Fig. \ref{Fig1}). Thus to get the density of states, one has
to sum the energies of these states over all values of $\gamma_2$: 
\begin{eqnarray}
 \rho(E) = \sum_{\gamma_2 = -\pi/2}^{\pi/2} \delta(E - E_{bound})
 \label{DOS}
\end{eqnarray}
and talk, in fact, about Andreev bands rather that single states. However, all
that was said on properties of the bound states, remains true for Andreev 
bands too. In particular the splitting of the whole band due to the spontaneous 
current is illustrated in the Fig. \ref{Fig3}.
The additional structure comes from the other (higher order) Andreev 
reflections. Superconducting energy gap $\Delta_0 = 0.376$ in this figure.

There is also a strong correlation between Andreev bound states (bands) and 
the pairing amplitude \cite{Vecino,KGA,KGA_1}. Each time the pairing amplitude 
at the $I$/$FM$ interface changes its sign, the Andreev bound state (band)
crosses the Fermi energy. Moreover in this case the spontaneous current is
generated. 

\begin{figure}[h]
 \resizebox{9cm}{!}{
  \includegraphics{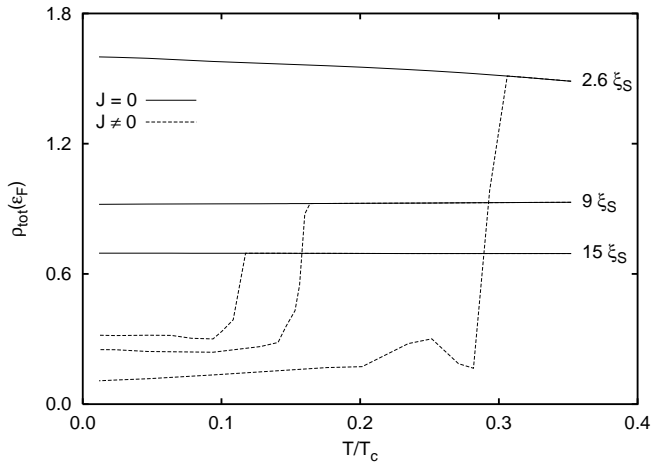}}
 \caption{\label{Fig4} The temperature dependence of the surface ($FM$/vacuum) 
          density of states at the Fermi energy for various thicknesses of the 
	  $FM$ slab in the figure. The solid (dashed) line corresponds to the 
	  solution without (with) the current. From Ref. \cite{KGA_2}}
\end{figure}

From the experimental point of view the density of states, in particular its
temperature dependence, can be a good measure of the current carrying ground
state. At certain thicknesses of $FM$ for which the current flows there is a 
huge drop in the $\rho_{tot}(\varepsilon_F)$ at characteristic temperature 
$T^{\ast} \approx (\xi_S/\lambda) T_c$, where $\xi_S$ and $\lambda$ are 
coherence length and penetration depth respectively. $T^{\ast}$ simply 
indicates the temperature at which magnetic instability, which leads to the 
generation of the current, takes the place. Such behavior is depicted in the
Fig. \ref{Fig4} and should be observable experimentally. If there is no current
the $DOS$ is due to the Andreev band and is almost constant (we are well below 
$T_c$), and as soon as the current starts to flow the Andreev band splits so we
observe a drop in $\rho_{tot}(\varepsilon_F)$. The important point is that 
$T^{\ast}$ and $T_c$ are different temperatures.
%


\section{\label{sec4} Spontaneous current}

\begin{figure}[h]
 \resizebox{9cm}{!}{
  \includegraphics{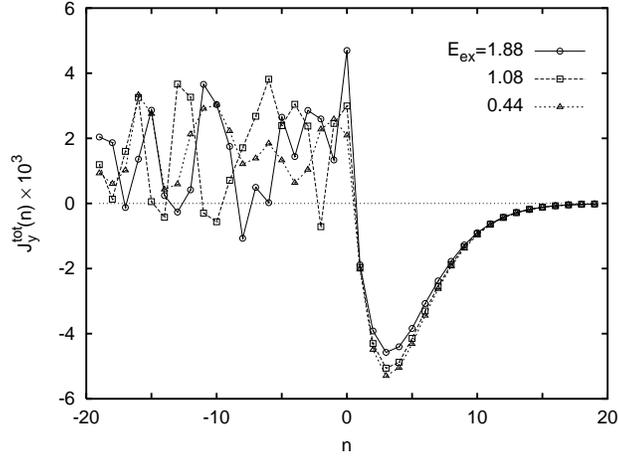}}
 \caption{\label{Fig5} The total (spontaneous) current
 $j^{tot}_y(n) = j_{y\uparrow}(n) + j_{y\downarrow}(n)$ flowing parallel to 
 the $FM$/$SC$ interface for a number of exchange splittings. From 
 Ref. \cite{KGA_1}.}
\end{figure}

The most remarkable feature of our calculations is that the solution of the 
$SPHFG$ equations frequently converges to a solution with the finite current 
$j_y(n)$ even though the external vector potential is zero. The typical example 
of such a current, flowing parallel to the $FM$/$SC$ interface, 
($j^{tot}_y(n) = j_{y\uparrow}(n) + j_{y\downarrow}(n)$) is shown  in the 
Fig. \ref{Fig5} for a few values of the exchange splitting. 
Behavior of the current, as a function of the layer index, is very similar to 
the density of states at the Fermi level. The oscillating nature of the current 
comes from the Friedel like oscillations of the $DOS$ \cite{KGA_1}. This is 
because current is proportional to the $DOS$ at the Fermi level. Within
semiclassical calculations, which neglect these effects the current is very
smooth \cite{KGA_3}.

\begin{figure}[h]
 \resizebox{9cm}{!}{
  \includegraphics{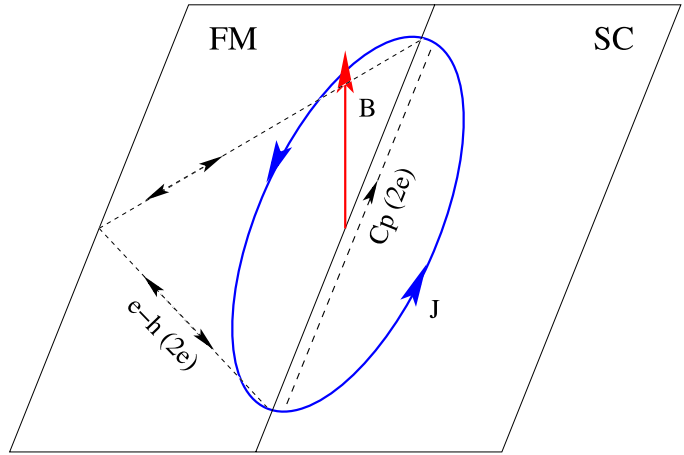}}
 \caption{\label{Fig6} Schematic view of the current distribution.}
\end{figure}

Another important issue is the distribution of the current through the whole
trilayer structure. We find that it flows mostly in the positive $y$ direction 
on ferromagnetic side and in the negative direction in the superconductor.
Notably the total current, integrated over the whole sample, is equal to zero 
within numerical accuracy. This is as it should be for the true ground state 
and found to be in the $FFLO$ state, where the current associated with the 
unpaired electrons is balanced by the supercurrent flowing in the opposite 
direction. Similarly here (see Fig. \ref{Fig6}).

\begin{figure}[h]
 \resizebox{9cm}{!}{
  \includegraphics{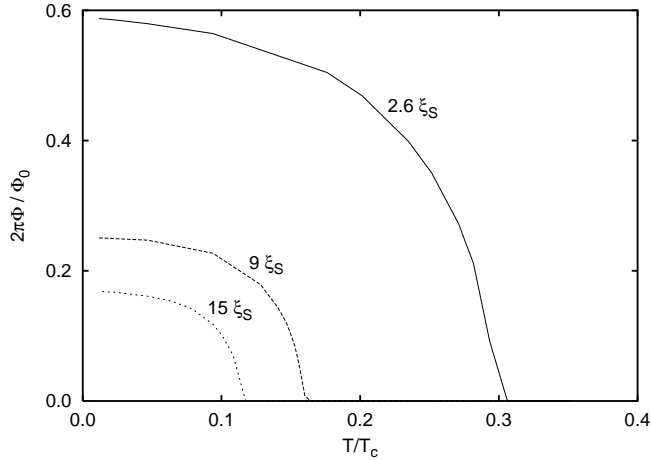}}
 \caption{\label{Fig7} The temperature dependence of the total magnetic flux 
          for thickness of the $FM$ slab $L/\xi_S = 2.6$ (solid), $6$ (dashed) 
	  and $15$ (dotted curve).}
\end{figure}

Obviously, the spontaneous current distribution (see Fig. \ref{Fig6}) 
generates the magnetic field through the sample. The total magnetic flux 
weakly depends on the thickness of the sample and the exchange splitting. Its
magnitude is found to be a fraction of the flux quantum $\Phi_0 = h/2 e$ and 
is smaller than upper critical field of the bulk superconductor. This is rather
a large field and could be observable in temperature dependent measurements 
(see Fig. \ref{Fig7}).
%


\section{\label{sec5} Transport properties}


Some information on spontaneous currents can be also obtained from conductance
calculations. To do so we attached a normal metal electrode to our $FM$/$SC$ 
system and calculate current through $NM$/$FM$/$SC$ system using nonequilibrium 
Keldysh Green's function technique \cite{Keldysh}. To calculate this current in 
terms of various physical processes we went along the way outlined in 
Ref. \cite{Cuevas} and got corresponding spin polarized formula for the current 
as a sum of four different contributions $I = I_1 + I_2 + I_3 + I_A$, where:
\begin{eqnarray}
I_1 = 4 \pi^2 t^2_{NF} \frac{e}{\hbar} \sum_{\sigma} \int d\omega
| 1 + G^{11r}_{FN\sigma}(\omega) |^2
\rho^{11}_{NN\sigma}(\omega) \rho^{11}_{FF\sigma}(\omega)
[f(\omega-eV) - f(\omega)]
 \label{I_1}
\end{eqnarray}
\begin{eqnarray}
I_2 = 8 \pi^2 t^2_{NF} \frac{e}{\hbar} \sum_{\sigma} \int d\omega
{\rm Re}\{ t_{NF} G^{21a}_{NF\sigma}(\omega)
[ 1 + G^{11r}_{FN\sigma}(\omega) ] \}
\rho^{11}_{NN\sigma}(\omega) \rho^{12}_{FF\sigma}(\omega)
[f(\omega) - f(\omega-eV)]
 \label{I_2}
\end{eqnarray}
\begin{eqnarray}
I_3 = 4 \pi^2 t^4_{NF} \frac{e}{\hbar} \sum_{\sigma} \int d\omega
| G^{12}_{FN\sigma}(\omega) |^2
\rho^{11}_{NN\sigma}(\omega) \rho^{22}_{FF-\sigma}(\omega)
[f(\omega-eV) - f(\omega)]
 \label{I_3}
\end{eqnarray}
\begin{eqnarray}
I_A = 4 \pi^2 t^4_{NF} \frac{e}{\hbar} \sum_{\sigma} \int d\omega
| G^{12}_{FF\sigma}(\omega) |^2
\rho^{11}_{NN\sigma}(\omega) \rho^{22}_{LL-\sigma}(\omega)
[f(\omega-eV) - f(\omega+eV)]
 \label{I_A}
\end{eqnarray}

$I_1$ corresponds to normal electron tunneling between electrodes, $I_2$ is a
net transfer of single electron with creation or annihilation of pairs as an
intermediate state. $I_3$ corresponds to a process in which electron from normal
electron is converted to a hole in superconductor - branch crossing process in
language of $BTK$ theroy \cite{BTK}, while $I_A$ is the Andreev tunneling. 

\begin{figure}[h]
 \resizebox{9cm}{!}{
  \includegraphics{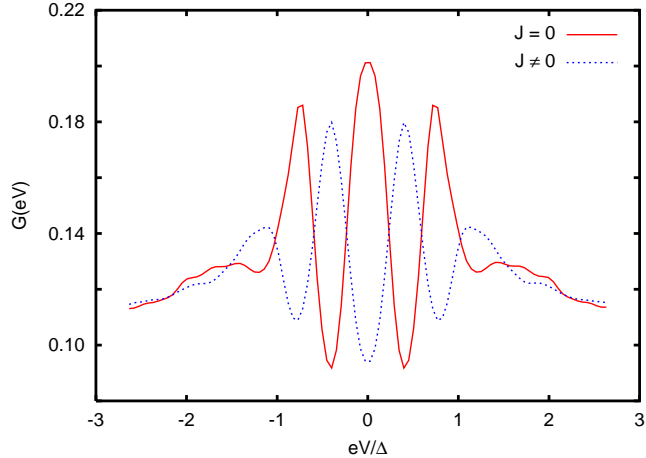}}
 \caption{\label{Fig8} The total differential conductance for the solution with 
          and without the spontaneous current.}
\end{figure}

The differential conductance $G(eV) = dI/d(eV)$ as a function of 
$eV = \mu_{NM} - \mu_{SC}$ is shown in the Fig. \ref{Fig8}.
Clearly, if there is a spontaneous current in the ground state, the conductance
peak is split, similarly as in the $DOS$. We could expect such behavior because
$G(eV)$ is proportional to the $DOS$ at the Fermi energy. And again this effect
could be observable in the tunneling experiments.

\begin{figure}[h]
 \resizebox{9cm}{!}{
  \includegraphics{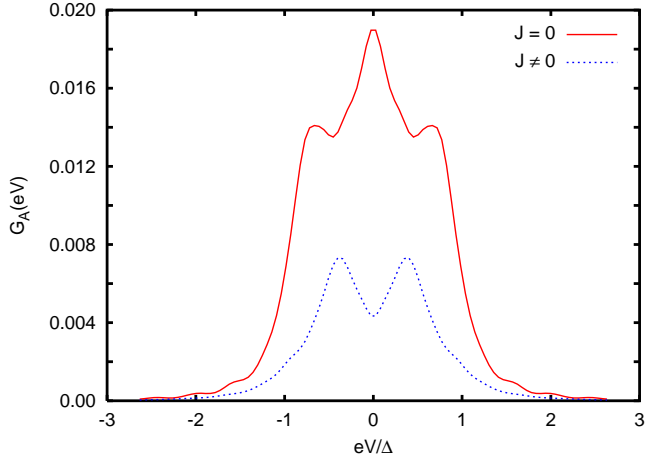}}
 \caption{\label{Fig9} Corresponding Andreev differential conductance for the 
          solution with and without the spontaneous current.}
\end{figure}

We have also extracted Andreev conductance form the total one and ploted in the
Fig. \ref{Fig9}.
We can see that conductance associated with the Andreev processes is strongly
enhanced when the current flows in the ground state. Unfortunately it could be
very difficult experimentally measure Andreev conductance only. Despite the
fact that for energies less than $SC$ gap the only allowed process is Andreev
reflection, as in the point contact geometry, it doesn't work in our system. 
The problem is that even at very low energies there is a finite $DOS$ at the 
Fermi level due to ferromagnet. Naturally the pairing amplitude is induced in 
$FM$ slab but this is not true energy gap in the quasiparticle spectra and we 
always deal with some single electron processes in tunneling events.


\section{\label{sec6} 2D FFLO state}


Before closing discussion on the spontaneous current we wish to make a remark
regarding the nature of the ground state in our system. To begin with we recall
that recently it has been predicted \cite{Izyumov} that under certain 
conditions a $3 D$-$FFLO$ state is energetically more favorable than usual 
$1 D$ state. The $3 D$ state manifests itself in oscillatory behavior of the 
pairing amplitude not only in the direction perpendicular to the interface but 
also in direction parallel to it. Moreover, changing the thickness of the $FM$ 
slab, one can switch the ground state of the system between $3 D$ and 
$1 D$-$FFLO$ state \cite{Izyumov,KGA_1}. 

The current carrying ground state of our system can be interpreted as a 
$2 D$-$FFLO$ state. The argument is as follows: The oscillations of the pairing 
amplitude in the direction perpendicular to the interface occur regardless 
whether the spontaneous current flows or not. Within the $FFLO$ theory 
\cite{FF,LO}, the period of the oscillations is related to the $x$-component of 
the center of mass momentum of the Cooper pair ${\bf Q}$. On the $FM$ side of 
our model the $FFLO$ periodicity is governed by 
${\bf Q} = (2 E_{ex} / v_F) \frac{{\bf v}_F}{v_F}$, where ${\bf v}_F$ is the 
Fermi velocity vector. This can be interpreted as the usual $1 D$-$FFLO$ state 
in confined geometry. On the other hand, when the current flows parallel to the 
interface, there is a finite vector potential in the $y$-direction. This can be 
regarded as a $y$-component of the ${\bf Q}$-vector. So one can say that when 
the spontaneous current flows, the $2 D$-$FFLO$ state is realized. Moreover 
when the $FM$ thickness is changed the ground state of the system is switched 
between $2 D$- and $1 D$-state, which manifests itself in spontaneous current 
flow or in the lack of it. In the present calculations this vector was found 
during the self-consistency procedure, as it is related to the vector potential 
in the $y$-direction. Moreover, the effective $Q_y$ changes its value from 
layer to layer leading to inhomogeneous $FFLO$-like state in both dimensions.


\section{\label{sec7} Conclusions}


The competition between ferromagnetism and superconductivity in $FM$/$SC$
heterostructures give raise to the Fulde - Ferrell - Larkin - Ovchinnikov
($FFLO$) state in these systems. The original bulk $FFLO$ state manifests 
itself in a spatial oscillations of the $SC$ order parameter as well as in 
spontaneously generated currents flowing in the ground state of the system. 
We have argued that a very interesting version of this phenomenon accures in 
$FM$/$SC$ proximity systems. In short, due to the proximity effect and Andreev 
reflections at the $FM$/$SC$ interface, the Andreev bound states appear in the 
quasiparticle spectrum. These states can be shifted to the zero energy by 
tuning the exchange splitting or the thickness of the ferromagnet, thus they 
became zero-energy mid-gap states which lead to various interesting effects. In 
particular, the occurence of spontaneous currents in the ground state can be 
related to the zero-energy states, as in the case of high-$T_c$ 
superconductors. It seems that some combination of both phenomena is realized 
in a real systems. The fact that oscillatory behavior of $SC$ order parameter 
is strongly correlated with the crossing of the Andreev bound states through 
Fermi energy and the generation of the spontaneous currents further support 
$FFLO$ - Andreev bound states picture. The experimental confirmation of the 
existence of the spontaneous (spin polarized) currents in the ground state 
would support the $FFLO$ - Andreev bound states scenario in these structures.

{\bf Acknowledgements:} This work has been partially supported by the grant 
no. PBZ-MIN-008/P03/2003. \\
BLG would like to thank the Center for Computational Material Science (CMS) of 
TU Wien for hospitality during the preparation of the above talk.



\end{document}